\documentclass[prb,aps,twocolumn,dcolumn,superscriptaddress,showpacs,amsmath,amssymb,showkeys]{revtex4}

\usepackage{graphicx,subfigure}
\usepackage{psfrag}
\usepackage{amsmath, amssymb,amsfonts}
\usepackage{siunitx}
\usepackage{subfig}
\usepackage{empheq}
\usepackage{filecontents}
\usepackage{float}
\usepackage[normalem]{ulem}
\usepackage[font=small,justification=raggedright,singlelinecheck=false]{caption}
\def\lto{Li$_{\rm 1+x}$Ti$_2$O$_4$}

\def\bE{\mbox{\boldmath$ E$}}

\def\bn{\mbox{\boldmath$ n$}}

\def\bu{\mbox{\boldmath$ u$}}

\begin{document}
\title{Multi-physics simulations of lithiation-induced stress in \lto\ electrode particles}
\author{Tonghu Jiang}
\affiliation{Department of Materials Science and Engineering, Johns Hopkins University, Baltimore, MD 21218 USA}
\author{Shiva Rudraraju}
\affiliation{Department of Mechanical Engineering, University of Michigan, Ann Arbor, MI 48109 USA}
\author{Anindya Roy}
\affiliation{Department of Materials Science and Engineering, Johns Hopkins University, Baltimore, MD 21218 USA}
\author{Anton Van der Ven}
%\affiliation{Department of Materials Science and Engineering, University of Michigan, Ann Arbor, MI 48109 USA}
\affiliation{Materials Department, University of California, Santa Barbara, CA 93106 USA}
\author{Krishna Garikipati}
\affiliation{Department of Mechanical Engineering, University of Michigan, Ann Arbor, MI 48109 USA}
\affiliation{Department of Mathematics, University of Michigan, Ann Arbor, MI 48109 USA}
\author{Michael L. Falk}
\email[Corresponding author. E-mail address: ] {mfalk@jhu.edu}
\affiliation{Department of Materials Science and Engineering, Johns Hopkins University, Baltimore, MD 21218 USA}
\affiliation{Department of Mechanical Engineering and Department of
  Physics and Astronomy, Johns Hopkins University, Baltimore, MD 21218
  USA}

\begin{abstract}    
Cubic spinel \lto\ is a promising electrode material as it exhibits a high lithium diffusivity and undergoes minimal changes in lattice parameters during lithiation and delithiation, thereby ensuring favorable cycleability. The present work is a multi-physics and multi-scale study of \lto\ that combines first principles computations of thermodynamic and kinetic properties with continuum scale modeling of lithiation-delithiation kinetics. Density functional theory calculations and statistical mechanics methods are used to calculate lattice parameters, elastic coefficients, thermodynamic potentials, migration barriers and Li diffusion coefficients. These quantities then inform a phase field framework to model the coupled chemo-mechanical evolution of electrode particles. Several case studies accounting for either homogeneous or heterogeneous nucleation are considered to explore the temporal evolution of maximum principle stress values, which serve to indicate stress localization and the potential for crack initiation, during lithiation and delithiation.

\end{abstract}
\maketitle

%\maketitle

\section{Introduction}
Scientists and engineers are exploring a vast materials space for cheaper, higher energy-density, and reliable battery materials, in response to the growing demand of portable electronics and electric vehicles.\cite{Hautier2011} Understanding the suitability of the cathode and anode materials, both computationally and experimentally, has been central to battery research. Li-ion batteries are the most used rechargeable batteries today, and a considerable fraction of battery research is focused on improving the characteristics of Li-ion batteries and finding better candidates for the electrodes.
 
In this work, we are interested in \lto\ for its possible use as an electrode material. \lto\ is known to form different polymorphs upon lithium insertion, and the common phases include rutile, anatase, spinel, and brookite.\cite{Cava1984,Zachau1988} The electric potential for lithium insertion into these structures is relatively low ($\sim$1.5V ), enhancing the suitability of \lto\ as an anode material. Cubic spinel \lto\ and Li$_{4/3}$Ti$_{5/3}$O$_4$ have been shown in experiments to exhibit high lithium diffusivity, and a negligible change of lattice parameter during (dis)charging.\cite{Krtil2001,Colbow1989,Wang1999} A large change in stress during the charging process can result, via the elastic coefficients, from significant changes in lattice parameters. The high stress can initiate fracture, limiting cycleability. Low variations in lattice parameters can thus promote better cycleability.

This work presents a multi-physics and multi-scale case-study in which we investigated \lto\ computationally, with the aim of predicting, from first principles, the stresses that will arise within the material during repeated charging and discharging of electrode particles. In particular, we are interested in spinel \lto, which belongs to the space group $Fd\bar{3}m$. In spinel \lto, 16$d$ and 32$e$ sites are occupied by Ti and O, respectively. We emphasize that the goal of the present work is to demonstrate how the various computational techniques at different length scales could be applied in tandem, and we used a high temperature of 800~K to magnify the kinetic effects. While specific quantitative predictions regarding electrode failure are not possible without continued development of these methods, the work presented here demonstrates that interesting insights into the chemo-mechanical processes that can limit cycleability can nonetheless be obtained.

\begin{figure}
\includegraphics[width=2.6in]{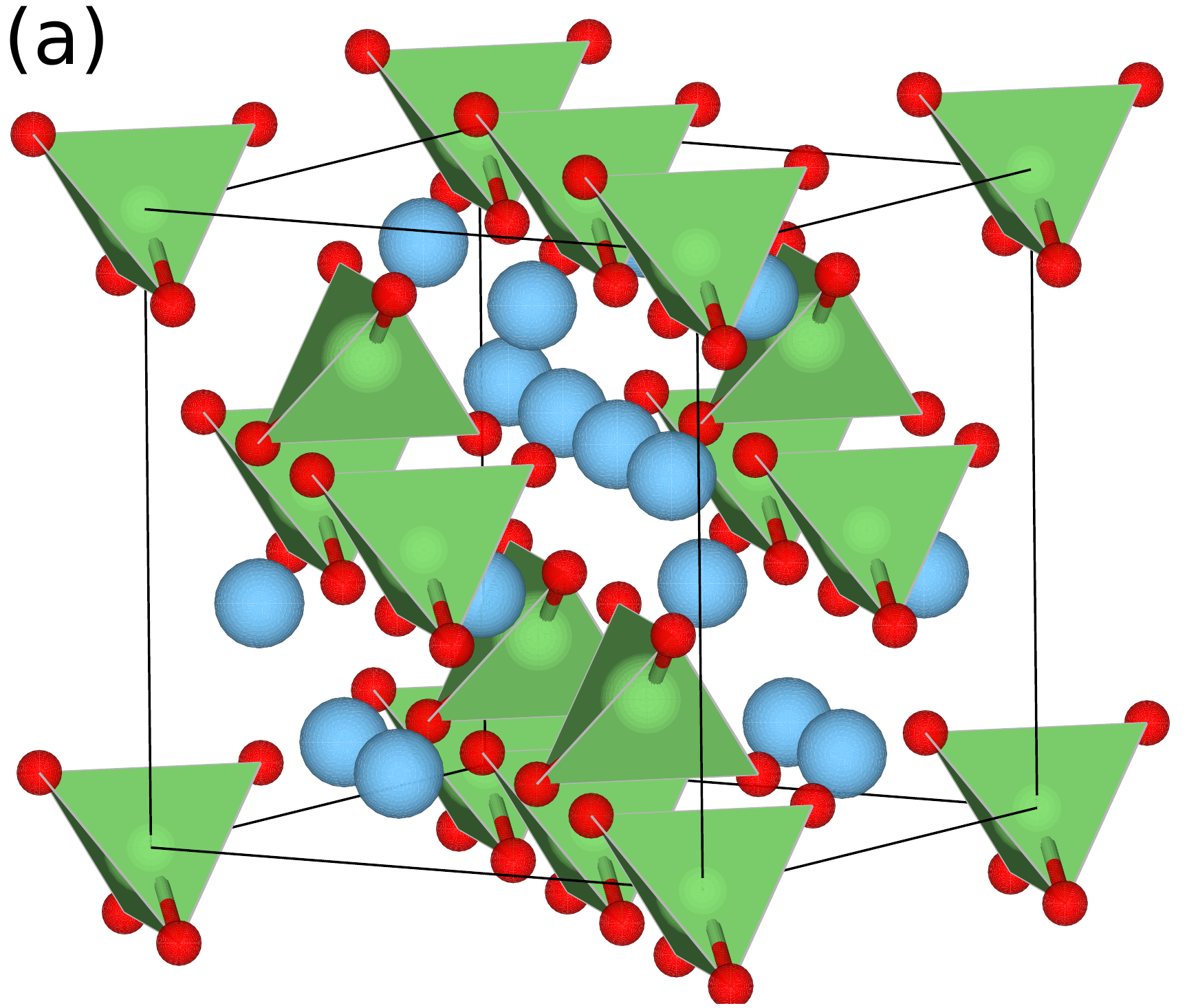} 
\includegraphics[width=2.5in]{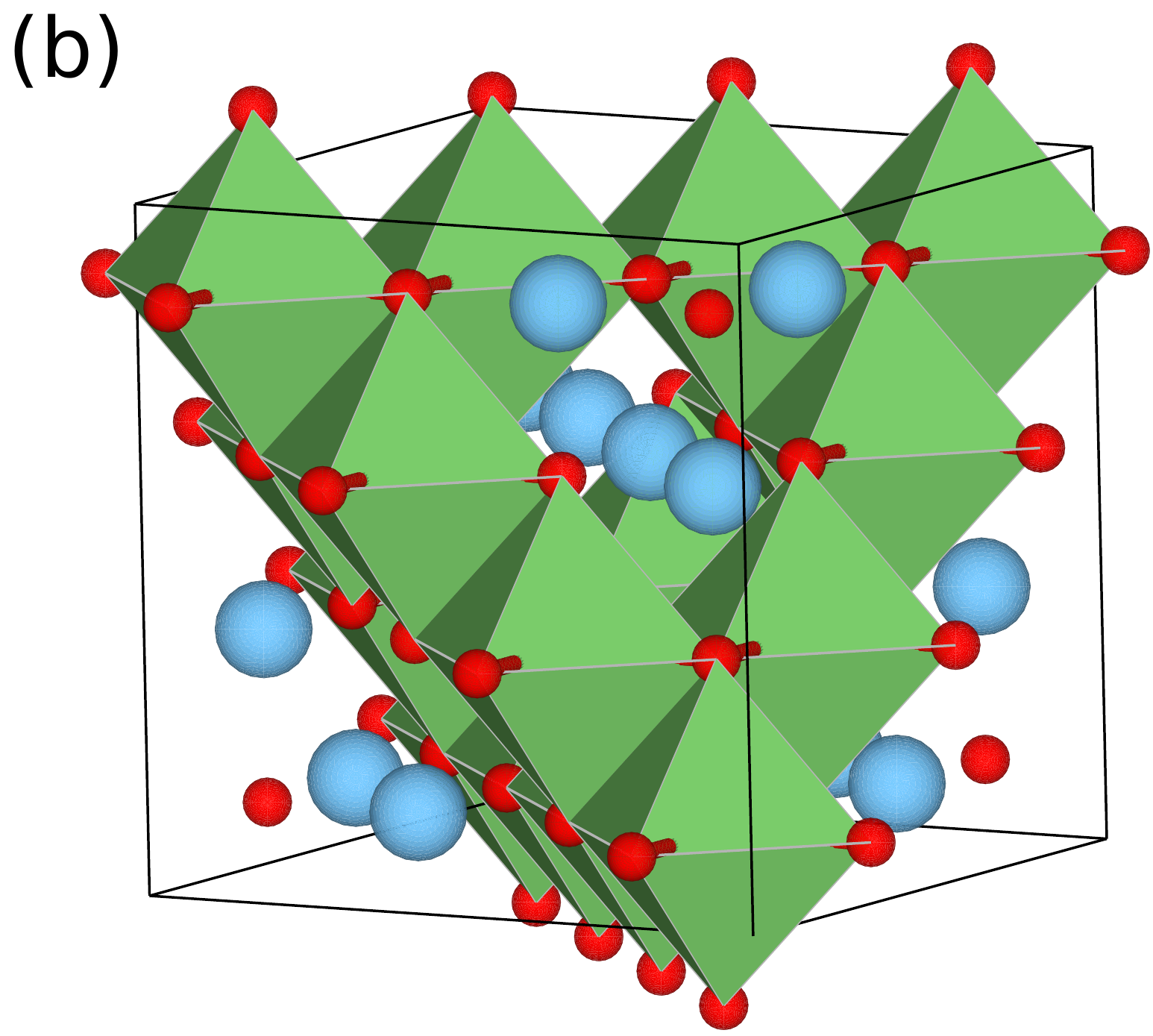} 
\includegraphics[width=2.4in]{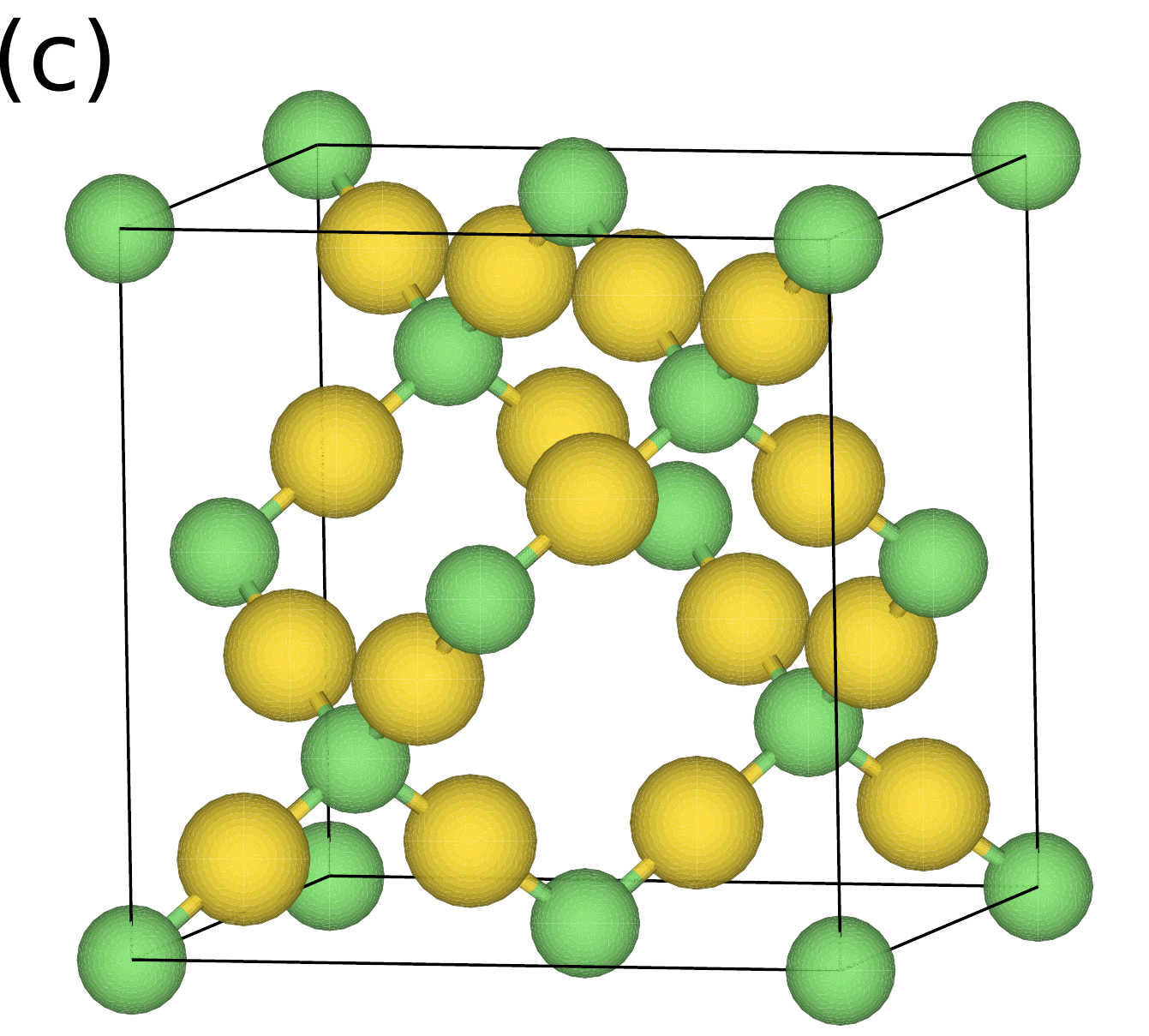} 
\caption{\label{fig1}  (a) and (b): The  ideal structure of $\alpha$ phase Li$_1$Ti$_2$O$_4$ and $\beta$ phase Li$_2$Ti$_2$O$_4$, respectively. The lithium ions reside in the tetrahedral sites and the octahedral sites formed by oxygen anions in respective Li$_1$Ti$_2$O$_4$ and Li$_2$Ti$_2$O$_4$. (c) Connection between lithium sites. The tetrahedral sites are represented by larger atoms than octahedral sites.  }
\end{figure}

During charge and discharge, \lto\ undergoes first order phase transformations. For x$<=$0, \lto\ is a solid solution $\alpha$ phase, in which the tetrahedral 8$a$ sites are occupied by lithium ions; after lithium insertion beyond x = 0, a second $\beta$ phase forms, in which lithium ions reside in octahedral 16$c$ sites. Between x = 0 and 1, lithium intercalation in \lto\ involves a two-phase coexistence process. Figure~\ref{fig1} shows the structures of the ideal $\alpha$ and $\beta$ phases in (a) and (b). The thermodynamic and electronic properties of  \lto\ have been studied by experiments and first-principles calculations.\cite{Krtil2001,Wagemaker2005,Bhattacharya2010} 

Lithium diffusion mechanisms and phase boundary mobility strongly influence overall battery performance. 
%For example, computational studies and experimental observations revealed that the phase transformation within LiFePO$_4$ at the atomic scale is dominated by one-dimensional lithium diffusion\cite{Ouyang2004} and interfacial strain. This eventually led to the development of a battery with a superfast (dis)charge rate.\cite{Kang2009} Furthermore, the strains that develop due to the phase transformation are responsible for limiting cycleability of the electrode material. Lithium-ion transport is reported to be limited by the interface mobility and lithium diffusion during the two-phase process.\cite{Li2012, bae1996, Funabiki1999}
%
Important factors that affect macroscopic lithium insertion and removal during a topotactic two-phase reaction within a crystalline electrode are interfacial free energies and coherency strain energy.  Lithium ion diffusion within the interface region must sustain the propagation of the phase boundary in a direction that will lower the overall free energy.   Lithium ion diffusion in most solid solution phases of electrode materials is anisotropic due to the crystal structure. However, because of its cubic symmetry, lithium diffuses isotropically in the solid solution phase of the spinel \lto\ . This is in contrast to two-dimensional diffusion in layered LiCoO$_2$\cite{anton2001} and one-dimensional diffusion in LiFePO$_4$.\cite{Ouyang2004, Morgan2004} 
As for the coherency strain energy of two-phase coexistence, the change in the lattice parameter is observed to be less than 1\% between the $\alpha$ and $\beta$ phases of \lto, and hence may not induce significant deformation of the interface. It is therefore a reasonable approximation to calculate homogeneous free energies at equilibrium lattice parameters from first principles and combine these with the {\em ab initio} computed change in lattice parameters and elastic moduli in continuum scale simulations to study the coupling between mechanics and chemistry during intercalation and deintercalation.

We have used an array of computational tools to study the thermodynamics and  kinetics of \lto.  Density functional theory computations were used to obtain formation energies of \lto\ with different lithium configurations and migration barriers for lithium ion hops. The cluster expansion method was employed to parameterize the formation energy and migration barrier dependence on Li-vacancy disorder within \lto. Cluster expansions and the Metropolis Monte Carlo (MMC) method were used to determine thermodynamic potentials, such as the composition-dependent Gibbs free energy and interfacial free energies. Mobility and diffusion in the solid solution phase were estimated using a combination of first-principles and kinetic Monte Carlo (KMC) computations.

Informed by quantum-mechanics- and statistical-mechanics computations, we then performed continuum scale computations for Li-ion diffusion and the mechanical stress induced by phase transformations in a phase field framework. The continuum scale computations used the following inputs from first principles computations: the free energies of the two stable phases, phase kinetics, interfacial energy and  changes in the lattice parameters and  elastic moduli. These computations are helpful in understanding the charge-discharge (lithiation-delithiation) kinetics and mechanical deformation that can induce mechanical failure in the anode particles. We studied the effects of different charge-discharge cycles on phase boundary movement, charge localization and peak stress evolution.

In the next section we elaborate the computational methods used in this work. Sec.~\ref{results} contains the main results of this work -- the multi-physics approach applied to the study Li-ion kinetics in \lto. 

\section{Methods}
\label{meth}

In this section we describe the computational methods in this multi-physics approach, from atomistic first principles calculations to the continuum scale formulation. The thermodynamics and kinetics of the $\alpha$ and $\beta$ phases of \lto\ have been previously studied by Bhattacharya {\it et al.}\cite{Bhattacharya2010} In this work, we use  first principles computations of the thermodynamics, kinetics and elastic constants to inform  continuum scale computations. This includes the composition dependence of lattice parameters and elastic coefficients. The continuum scale computations use a coupled phase field and finite strain mechanics multi-physics formulation. The use of the finite strain (nonlinear elasticity) formulation distinguishes this work, as phase field computations are traditionally limited to the infinitesimal strain approximation and thus only use linear elasticity formulations. In Sections \ref{DFT}--\ref{mcc} we discuss the first-principles methods, followed by the thermodynamics and the Li-ion kinetics calculated for the \lto\ system. In Section \ref{subsec:phasefieldmodel}, we elaborate on the continuum scale formulation.

\subsection{First principles calculations}
\label{DFT}

We constructed a first-principles cluster expansion to describe the energy of \lto\ as a function of Li-vacancy order/disorder. To accomplish this, we borrowed formation energies of 48 Li-vacancy orderings in small supercells of the primitive \lto\ unit cell and 4 migration barriers from the work of Bhattacharya {\it et al.},\cite{Bhattacharya2010} and extended this dataset by adding more energy information sampled in a larger configuration space. We used the VASP\cite{Kresse1996a,Kresse1996b}  plane wave code to relax structures with different lithium configurations and calculate their total energies. The parameters used in the calculations are referred to in Ref.~\citenum{Bhattacharya2010}.  

The lattice parameters of \lto\ can be extracted from the relaxed structures. Because the cell shape does not necessarily remain cubic after relaxation, we calculated the volumes of the relaxed structures and then converted them to lattice parameters assuming the supercells were cubic. The elastic constants are key parameters for studying the intercalation-induced mechanical response of Li$_{\rm 1+x}$Ti$_2$O$_4$. Because off-stoichiometric Li concentrations in \lto usually have arrangements that break the cubic symmetry of ideal Li$_{1}$Ti$_2$O$_4$ or Li$_{2}$Ti$_2$O$_4$, it becomes impractical to deform the structures and extract elastic constants from the energy-strain relationship. We calculated elastic constants from the stress-strain relationship using finite-differences, obtained using VASP.\cite{Page2002} To calculate the composition dependence of elastic constants, we followed the method by Liu {\it et al.}\cite{Liu2005prb} to symmetrize the elastic constants by transforming the original elastic constants by the 48 symmetry operations belonging to the cubic $m\bar{3}m$ point group.

The migration barriers of lithium were calculated with the nudged elastic band method implemented in VASP. We identified 18 symmetrically distinct Li hops between pairs of stable tetrahedral and octahedral sites over a range of Li concentrations in addition to the four lithium hops considered by Bhattacharya {\it et al.} \cite{Bhattacharya2010}. The migration barriers were calculated in a conventional spinel cell. 

%To calculate the migration barriers for lithium in different lithium configurations, particularly in the two-phase region, we first relaxed more than 100 structures with random lithium ion distributions in a conventional cell to obtain stable structures. For each stable structure we moved one lithium ion at a time from an octahedral  (tetrahedral) site to an adjacent tetrahedral (octahedral) site, and then relaxed the new structure to determine if every lithium ion in the new structure was stable. All of the stable pairs of structures were compared with each other under local symmetry operations to remove redundancy. In this way we identified 17 lithium transitions in addition to Bhattacharya's 4 lithium transitions. 

\subsection{Cluster expansion}
\label{CE}

The cluster expansion method by Sanchez {\it et al.}\cite{Sanchez1984, Laks92, Fontaine94} is a mathematical tool to describe the configuration dependence of any property of a multi-component crystal and has contributed greatly to the development of modern alloy theory in oxides.\cite{Ceder2000} It has been successfully applied in the study of various crystalline materials, particularly when analyzing phase stability, order-disorder transitions and diffusion.\cite{Wolverton1998,anton2001} In this study, we are interested in cluster expanding formation energies and migration barriers. The 0~K formation energy of a particular Li-vacancy ordering (labeled $\sigma$) within \lto\ is defined  as:
\begin{equation}
\begin{aligned}
\Delta E\left(\sigma \right) &= E( {\rm Li_{1+x}Ti_2O_4}) \\
&- \frac{(1+x)E({\rm Li_2Ti_2O_4})+(1-x)E({\rm Ti_2O_4})}{2}
\end{aligned}
\end{equation}
where $E( {\rm Li_{1+x}Ti_2O_4})$ is the total energy of a Li-vacancy ordering $\sigma$ and where $E({\rm Li_2Ti_2O_4})$ and $E({\rm Ti_2O_4})$ are total energies of fully lithiated spinel (all octahedral sites occupied by Li) and delithiated spinel, respectively. The occupation of each lithium site, $i$, within \lto is represented by a site occupation variable $S_i$ that is +1 if occupied by a lithium atom and -1 if vacant. The Li-vacancy ordering $\sigma$ within \lto\ can then be uniquely determined by specifying all the occupation variables $S_{i}$. A cluster expansion expresses the formation energy $\Delta E (\sigma)$  as a sum of products of effective interactions and polynomials of occupation variables associated with clusters of sites according to \cite{Sanchez1984}:
\begin{equation}
\begin{split}
\Delta E (\sigma) = J_0 + \sum_{i}{J_iS_i(\sigma)} +
\sum_{j<i}{J_{ij}S_i(\sigma)S_j(\sigma)}&\\
+\sum_{k<j<i}{J_{ijk}S_i(\sigma)S_j(\sigma)S_k(\sigma)} +& \cdot\cdot\cdot
\label{eq1}
\end{split}
\end{equation}
The expansion coefficients ($J_{i}$, $J_{ij}$, etc.) are refered to as effective cluster interactions (ECI) and can be determined from first-principles using a variety of inversion methods that minimize a cross validation score \cite{Walle2002}. Models (i.e. selection of non-zero ECI in a truncated cluster expansion) can be generated with genetic algorithms \cite{Hart2005}, by deploying Bayesian approaches using prior knowledge,\cite{Mueller2009} or by implementing compressive sensing approaches.\cite{Nelson2013} We applied standard methodologies in our work, and the reader is referred to the cited references for further details.

\subsection{Thermodynamics and Kinetics}
\label{mcc}

First-principles calculations and cluster expansions together provide us with the composition dependence of lattice parameters, formation energies, and elastic coefficients at zero-temperature. The application of Monte Carlo simulations to the cluster expansion of the formation energy enables the calculation of the free energies of bulk phases and interfaces at finite temperature, as described below. Li-ion kinetics is also determined at finite temperature using kinetic Monte Carlo methods that rely on cluster expansions to describe the configuration dependence of migration barriers and the energies of the end states of Li hops. 

\subsubsection{Homogeneous free energy}

%The cluster expansion does not describe vibrational excitations or the elastic energy when the crystal is deformed away from its equilibrium lattice parameters. The Gibbs free energy calculated with data from Monte Carlo simulations applied to the cluster expansion therefore corresponds to the configurational free energy at ambient pressure. Vibrational contributions to the formation free energy can be neglected. 

%The Gibbs free energy is related to the grand thermodynamic potential $\Phi$ by $G=\Phi+\mu N$, where $\mu$ is the chemical potential, and we can calculate Gibbs free energies for the $\alpha$ and the $\beta$ phases by calculating $\Phi$ first. The calculation of free energy at finite temperature T$_0$ must start from a reference state whose free energy can be calculated exactly so that the free energies of the $\alpha$ and the $\beta$ phases are comparable. 

We employed thermodynamic integration, a commonly used free energy calculation technique, to calculate the grand thermodynamic potential $\Phi=U-TS-\mu N$,\cite{Walle2002,Binder2011,book3} which is related to the Gibbs free energy according to $G=\Phi+\mu N$ . At constant chemical potential the grand thermodynamic potential can be calculated using:
\begin{equation}
\frac{\Phi(T)}{k_bT}-\frac{\Phi(T_0)}{k_bT_0}=\int^T_{T_0} (E-\mu N) d\left(\frac{1}{k_bT}\right)
\label{integrate1}
\end{equation}
At constant temperature, the grand thermodynamic potential can be calculated with:
\begin{equation}
\Phi(\mu)-\Phi(\mu_0)=\int^{\mu}_{\mu_0}N(\mu)d\mu
\label{integrate2}
\end{equation}

\subsubsection{Interfacial free energy}

 Interfacial energy and surface energy play important roles when the size of the electrode crystallite approaches the nanoscale.  Following Binder,\cite{Binder2011} we calculated interfacial free energies, $F_{int}(T)$, by thermodynamic integration from $T_0$ to $T$ of the expression:
\begin{equation}
F_{int}(T)/kT=F_{int}(T_0)/kT_0+\int_{1/kT_0}^{1/kT} E^{ex}(T')d(1/kT')
\end{equation}
The $E^{ex}(T')$ is the excess energy due to the existence of an interface, defined by the following equation:
\begin{equation}
\label{excess}
E^{ex}(T)= E^{sys}(T)-x E_{\alpha}(T)-(1-x)E_{\beta}(T)
\end{equation}
Here $E^{sys}(T)$ is the total energy of the system of two coexisting phases, while $E_{\alpha}$ and $E_{\beta}$ are the total energies of pure $\alpha$ and $\beta$ phases of the same size as in the studied system, and $x$ is the fraction of $\alpha$ phase in the system.

\subsubsection{Lithium ion mobility calculation}

Mobility in a homogeneous crystalline solid is a quantity that describes collective transport of mobile atoms or ions. We first calculated the self-diffusion coefficient $D$ using kinetic Monte Carlo simulations to approximate the Kubo-Green expression\cite{Richards77}
\begin{equation}
\label{selfd}
D=\frac{\langle(\sum_{i}R_{i})^{2}\rangle}{2dNt}.
\end{equation}
In the above equation, $R_{i}$ is the displacement of the $i$-th Li ion after time $t$, $N$ is the number of Li ions, and $d$ is the dimension of the crystal. The self-diffusion coefficient $D$ measures the random walk process of the geometric center of mass of the mobile lithium ions. The lithium ion mobility can be obtained by converting $D$ through:
\begin{equation}
\label{M_D_relation}
M=\frac{x}{k_bT}D,
\end{equation}
where $x$ is the composition of the homogeneous system. 

\subsection{The phase field model} 
\label{subsec:phasefieldmodel}
Phase field models are commonly employed to study the evolution of spatially continuous order parameters by gradient flow kinetics. They have been widely adopted for modeling phase transformations described by the evolution of conserved (Cahn-Hilliard model \cite{Cahn1958}) or non-conserved order parameters (Allen-Cahn model\cite{Cahn1972}). In this work, the existence of stable $\alpha$ and $\beta$ phases at Lithium compositions of $x=0$~(Li$_1$Ti$_2$O$_4$) and $x=1$~(Li$_{2}$Ti$_2$O$_4$), respectively, and a two-phase co-existence at intermediate values of the composition allows for a classical double-well representation of the free energy as illustrated in Figure~\ref{fig:freeEnergy}. 
 
In modeling the diffusive kinetics and mechanical response of electrode particles, phase field treatments have traditionally taken into account the interfacial energy, elastic strain energy and anisotropies of the crystalline electrode material. Past studies used this approach to study the formation and growth of the solid-electrolyte interface layer,\cite{Deng2013} revealing a diffusion-limited process. Similarly, Kao {\it et al.} \cite{Kao2010} found via experiment and phase field modeling that the phase transition pathway depends on the overpotential in the lithium ion phosphate battery. Further, Lithium intercalation into nanometer-sized electrode particles,\cite{Burch2009} charge transport in TiO$_2$,\cite{Hu2013} and large mechanical stress due to phase segregation in Li$_{\rm x}$Mn$_2$O$_4$\cite{Huttin2012} have also been modeled using the phase field method.

In this work, we consider a phase-field model driven by anisotropic interfacial and elastic strain energies obtained from first principles computations. Traditionally, most phase field treatments coupled with mechanics were limited to the infinitesimal-strain assumption, which does not satisfy frame invariance of the elastic strain energy function. In the presence of large strains, this induces spuriously high stresses due to unaccounted rotations and renders the model susceptible to predicting failure by the wrong mechanisms. However, our treatment is framed within nonlinear kinematics (finite strains), which nullifies the effect of rotations on the elastic strain energy with mathematical exactness. As stated in the introduction, for spinel $\mathrm{Li}_{\rm 1+x}\mathrm{Ti}_2\mathrm{O}_4$, the diffusion of lithium is close to isotropic due to high symmetry of the lattice. We therefore consider an isotropic mobility for lithium-ion diffusion in the stable phases and the interface.

In this model, we consider the following free energy:
\begin{equation}
\Pi(x, \bE)=N_v\int[f(x) +\frac{1}{2}\kappa(\theta) (\nabla x)^2 + W(x, \bE)]dV
\label{freeEnergyEqn}
\end{equation}
where $x$ is the composition, $f$ is the homogeneous free energy density and $\bE$ is the Green-Lagrange strain tensor. The latter is a frame invariant measure of strain, whose components are given by $E_{IJ} = \frac{1}{2}(F_{kI}F_{kJ} - \delta_{IJ})$. Also in component form $F_{iJ} = \delta_{iJ} + \partial u_{i}/\partial X_{J}$ is the deformation gradient tensor and $u_i$ is the displacement vector. 

The first two terms in the integrand of Equation (\ref{freeEnergyEqn}) represent the homogenous free energy density and the orientation dependent gradient energy density of the interface, respectively. The form of the homogeneous free energy considered is shown in Figure~\ref{fig:freeEnergy} and the orientation dependence of the interfacial energy density is given by Figure~\ref{fig18}. Here, $\kappa$ is related to the interfacial energy as\cite{Cahn1958} 
\begin{equation}
\gamma = 2\int\limits_{x_\alpha}^{x_\beta}\sqrt{\kappa \varDelta f }\mathrm{d}x
\label{eq:intfcenergy}
\end{equation}
where $\varDelta f$ is the difference between homogeneous free energy density and the common tangent line between compositions at $x_\alpha$ and $x_\beta$ (Figure~\ref{fig:freeEnergy}). The exact form of homogeneous free energy density, $f$, is not very influential in the resulting phase microstructure or the dynamics by which it evolves.\cite{Vaithy2002} The magnitude of $\kappa$ is usually adjusted in order to reproduce the experimental or calculated interfacial free energy so long as the resulting interface width in the continuum scale computation is reasonable.

The third term in the integrand of Equation (\ref{freeEnergyEqn}) is the elastic strain energy density, which is a function of the Green-Lagrange strain and the composition. Assuming a St.~Venant-Kirchhoff type material model, the specific form of the elastic strain energy density considered here is given by:
\begin{equation}
W(x, \bE)=\frac{1}{2}(E_{IJ}-E^{x}(x)~\delta_{IJ})\mathbb{C}_{IJKL}(x)(E_{KL}-E^{x}(x)~\delta_{KL}) 
\label{strainEnergyEqn}
\end{equation}
where $\mathbb{C}$ is the fourth order elasticity tensor with major and minor symmetries and cubic anisotropy, and $E^{x}$ is the stress-free strain introduced to model changes in the lattice parameter as a function of the composition. The composition dependence of the elastic moduli and of the lattice parameter are obtained from first-principles computations described in the next section, and the composition dependence is depicted in Figure~\ref{fig:latt}. 

Using a variational approach, we can derive the governing equations for the dynamics of non-equilibrium chemistry and for mechanical equilibrium. The local chemical potential $\mu$ is obtained by evaluating the variational derivative of the free energy $\delta \Pi /\delta x$, and is given by
\begin{equation}
\mu=\frac{\partial f}{\partial x} - \kappa\nabla^2 x + \frac{\partial W}{\partial x}
\end{equation}
The conservation of mass governs the non-equilibrium chemistry, and is given by
\begin{equation}
\begin{aligned}
\label{dcdt}
\frac{\partial x}{\partial t} &= -\nabla (-M\nabla \mu)
\\
&=\nabla \Big[ M\nabla\big( \frac{\partial f}{\partial x} - \kappa\nabla^2 x + \frac{\partial W}{\partial x} \big)\Big]
\end{aligned}
\end{equation}
where $M$ is the constant mobility. This partial differential equation is complemented by boundary conditions:
\begin{equation}
\begin{aligned}
\nabla x\cdot\bn &= 0,\\
M\nabla\mu\cdot\bn &= j,
\end{aligned}
\label{eq:chembcs}
\end{equation}
where the first of the above boundary conditions is a consequence of assuming equilibrium at boundaries, while the second represents a boundary influx, $j$. Taking variations with respect to the displacement, we obtain the governing equation of quasi-static mechanical equilibrium (conservation of linear momentum), on the basis that elastic equilibrium is established much more rapidly than chemical equilibrium:
\begin{equation}
\frac{\delta W}{\delta \bu} = 0 \notag \\
\label{mechanics}
\end{equation}	
Following standard variational arguments this condition leads, via Euler-Lagrange equations, to the following partial differential equation:
\begin{equation}
P_{iJ,J}= 0 \\
\label{mechanics2}
\end{equation}	
\noindent where, $P_{iJ}=\frac{\partial W}{\partial F_{iJ}} $ are components of the stress tensor. The partial differential equation in (\ref{mechanics2}) is subject to boundary conditions:
\begin{equation}
\begin{aligned}
u_i = g_i\quad\mathrm{on} \;\Gamma_u\\
P_{iJ}N_J = T_i\quad\mathrm{on}\; \Gamma_P
\end{aligned}
\label{eq:mechbcs}
\end{equation}
where $\Gamma_u$ and $\Gamma_P$ are the Dirichlet and Neumann boundaries, respectively. In this formulation, the stress tensor inherits a composition dependence from the strain energy density $W(x, \bE)$.

Equations (\ref{dcdt}) and (\ref{mechanics2}) are the governing partial differential equations for non-equilibrium chemistry and for mechanical equilibrium, respectively. We 
use the Finite Element Method (FEM) to solve these partial differential equations in a weak (integral) formulation. The details of the weak formulation, finite dimensional discretization and the solution schemes adopted are beyond the scope of this paper, but the methods considered are standard in the FEM literature. The numerical framework to solve the system of equations was implemented in an in-house C++ FEM code built on top of the deal.II library,\cite{dealII} and uses the Sacado library of the Trilinos project \cite{Sacado} for algorithmic differentiation to generate the Jacobian matrix. Some of the details of the weak formulation, finite dimensional discretization and numerical framework can be found in related publications by the authors.\cite{Rudraraju2014, Rudraraju2016}
 
\section{Results and Discussion}
\label{results}

A significant fraction of the atomistic calculations needed to connect to the continuum scale calculations have been reported before by Bhattacharya {\it et al.} in Ref.~\citenum{Bhattacharya2010}, and we restrict our report to the results from the atomistic calculations that are new, or essential.

\subsection{Results of first-principles calculations}

\subsubsection{Lattice parameters and elastic constants}
\begin{figure}
\includegraphics[width=3.3in]{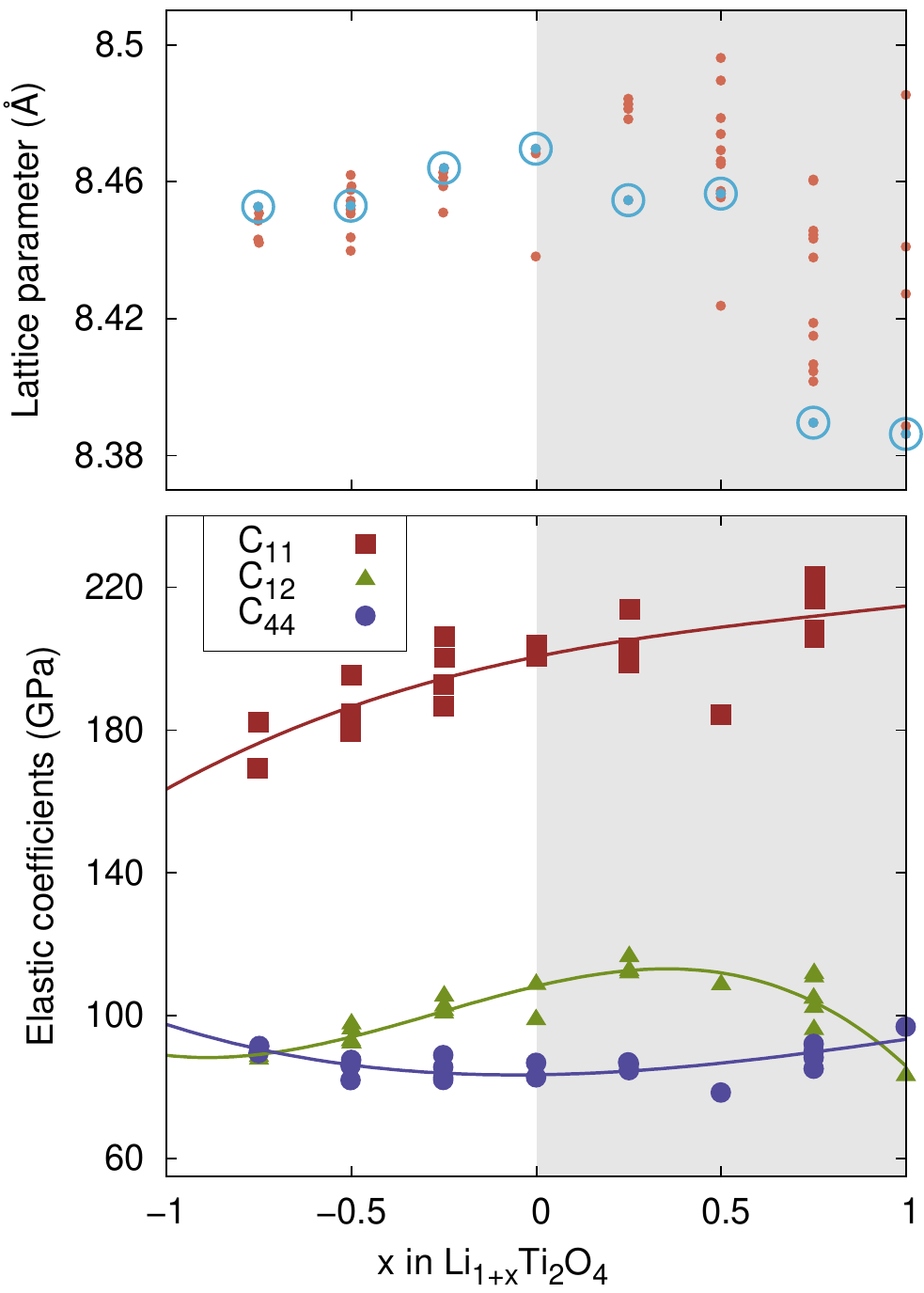} 
\caption{\label{fig:latt} (Top panel) lattice parameter of \lto\ calculated by DFT. Each red dot represents a lattice parameter of a unique structure. The dots in circle (in blue) represents the lattice parameter of the structure with the lowest energy. (Bottom panel) Red squares, green triangles and purple circles represent elastic coefficients C$_{11}$, C$_{12}$, and C$_{44}$ respectively. The solid lines through the group of points are third-order-polynomial fit to the calculated elastic coefficients.}
\end{figure}

Figure~\ref{fig:latt} (top panel) shows lattice parameters as a function of number of lithium ions in a conventional spinel cell. In the common phase field models the lattice parameter is assumed to be linearly dependent on composition following Vegard's law. In Figure~\ref{fig:latt} we see that the lattice parameter contracts after the phase transition from $\alpha$ to $\beta$, while in the $\alpha$ phase the lattice parameter increases with lithium ion density. Although the trend of lattice parameter is clear, the relative difference between Li$_{1}$Ti$_2$O$_4$ and Li$_{2}$Ti$_2$O$_4$ is small, i.e, $\sim$1\% for this phase transition.

Figure~\ref{fig:latt} (bottom panel) shows the lithium composition dependence of the elastic coefficients C$_{11}$, C$_{12}$ and C$_{44}$. The individual points correspond to calculated values, whereas the lines drawn through them are third order polynomial fits. As can be seen in Figure~\ref{fig:latt}, C$_{11}$ increases monotonically as the lithium composition increases. C$_{11}$ increases rapidly in the $\alpha$ solution phase from x$\approx$$-$1 to x$\approx$0 and slowly in the two phase region. The overall increase of C$_{11}$ from x=$-$1 to x=1 could be as large as 50 GPa or 30\% of its value at x$=-$1. C$_{11}$, C$_{12}$ and C$_{44}$ generally depend on lithium composition in a nonmonotonic way. The assumption made in many continuum scale models\cite{Hu2001} that elastic constants depend linearly on composition may not be satisfied particularly for C$_{12}$ and C$_{44}$. 

\subsection{Cluster expansions}

\subsubsection{Cluster expansion of \lto\ }

We are interested in building a cluster expansion capable of predicting energies in single phases and within interfaces. The lithium configurations at the interface region correspond to high energy states; thus they are not well described by cluster expansions that have been optimized to more accurately predict low energy configurations. Seko {\it et al.} proposed a procedure that improves the cluster expansion model by adding structure samples in a systematic way such that the correlation between the structures is reduced.\cite{Seko2009} In this work, we simply add more training configurations to the cluster expansion fit that have random lithium arrangements in a $1\times 1\times 1$ conventional cell. Thirty-six structures corresponding to the two end states of 18 lithium transitions in the NEB calculations were added in the cluster expansion fit. This improved the cluster expansion prediction of energy differences between the end states of Li hops, and therefore also led to a more accurate prediction of migration barriers, as discussed in Section~\ref{CE}. No more than 50 clusters were included in the cluster expansion fit to avoid overfitting, and a total of 92 formation energies of different configurations were used in the cluster expansion fit. The final cluster expansion includes 11 pair clusters, 9 triple clusters, and 3 quadruple clusters. The error and cross validation score of the cluster expansion is 18~meV and 27~meV per primitive cell.

To verify that this cluster expansion yields the correct phase behavior at 0~K, a series of Metropolis Monte Carlo calculations were performed that started at high temperature and that were slowly cooled down to 0~K to search for the lowest energy states. The lowest energy states obtained from Metropolis Monte Carlo calculation were then confirmed with a  genetic algorithm search. The results obtained from these calculations agree with the previous calculations reported by Bhattacharya in Ref.~\citenum{Bhattacharya2010}, and we refer the reader to Figure~2 therein and the associated text.

\subsection{Thermodynamics and Kinetics}

\subsubsection{Homogeneous free energy}
\begin{figure}[hbt]
%    \begin{tikzpicture}[scale=1.0]
%      \begin{axis}[minor tick num=1,xlabel={$\text{Li}_{1+\text{x}}\text{Ti}_2\text{O}_4$},ylabel={Free Energy (eV)},  xmin=-1, xmax=1.15,  ymin=-510, ymax=1000, legend entries={}, legend style={draw=none, at={(0.75,1)},anchor=north west}]
 %      	\addplot [black, thick] table [x={x}, y={y}] {\cFreeEnergy};
 %	\addplot [blue, only marks] table [x={x}, y={y}] {\cFreeEnergyFirstPrinciples};
 %     \end{axis}
 %   \end{tikzpicture}
    \includegraphics[width=2.5in]{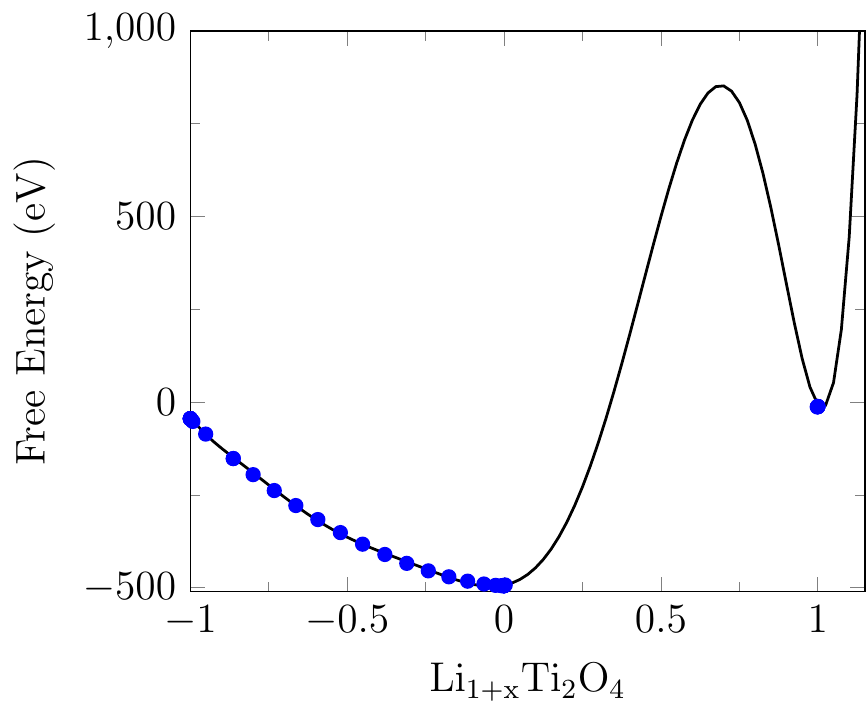} 
    \caption{\label{fig:freeEnergy}  Homogenous free energy representation obtained by fitting a double-well polynomial to the free energy values obtained from grand canonical Monte Carlo calculations. The blue dots are the free energy values obtained from grand canonical Monte Carlo calculations.}
\end{figure}

Figure~\ref{fig:freeEnergy} shows the calculated Gibbs free energies of the $\alpha$ and $\beta$ phases with Monte Carlo data obtained in a 4$\times$4$\times$4 supercell at T=800~K, denoted by blue dots. We can see that the $\beta$ phase is stable in only a very narrow lithium composition interval around $x \approx$1. The homogeneous free energy was fit with a double-well function (the solid black curve in Figure~\ref{fig:freeEnergy}) for use in the phase field model.

%\pgfplotstableread{data/freeEnergy.dat}{\cFreeEnergy}
%\pgfplotstableread{data/freeEnergyFirstPrinciples.dat}{\cFreeEnergyFirstPrinciples}

\subsubsection{Interfacial free energy}

%The interfacial energy strongly influences the equilibrium crystalline geometry. A thermodynamic analysis by Wagemaker {\it et al.}\cite{Wagemaker2009} illustrated how the interfacial free energy can alter the equilibrium concentrations of coexisting phases at the nanoscale, because of the increased importance of the interfacial free energy penalty as the size of the crystallite decreases. Moreover, anisotropic interfacial energy is essential in modeling microstructure evolution.\cite{Vaithy2002} It regulates the morphology of the precipitate together with strain energy. In most computational studies, due to the incoherent lattices of the two coexisting phases, the interfacial energy has to be calculated at T=0~K,\cite{Vaithy2002} or by Talyor expansion of the partition function at low temperature.\cite{Asta1998}

We constructed 8$\times$4$\times$4 and 10$\times$5$\times$5 Monte Carlo supercells to calculate finite temperature interfacial free energies (1$\times$1$\times$1 is a conventional cell). The periodic screw boundary condition was imposed on the system in the Monte Carlo computation. The implementation details of this boundary condition can be found in\cite{TonghuThesis}. A total of 15 nonequivalent orientations of the formula $(1,\frac{M}{4},\frac{N}{4})$, where M, N = 0,1,2,4, were selected in the 8$\times$4$\times$4 supercell, and similarly 21 orientations were selected for the 10$\times$5$\times$5 supercell. The Monte Carlo computations started out at 1200~K (below the order-disorder temperature) with sharp interfaces and were run for 15000 sweeps. The temperature was lowered by 50~K after every other sweep until the temperature reached 0~K. The final systems at 0~K are lower in energy than the initial systems with sharp interfaces. However, they are not guaranteed to be the lowest energy states because the final energy may be cooling-rate dependent.

\begin{figure} 
\includegraphics[width=3.5in]{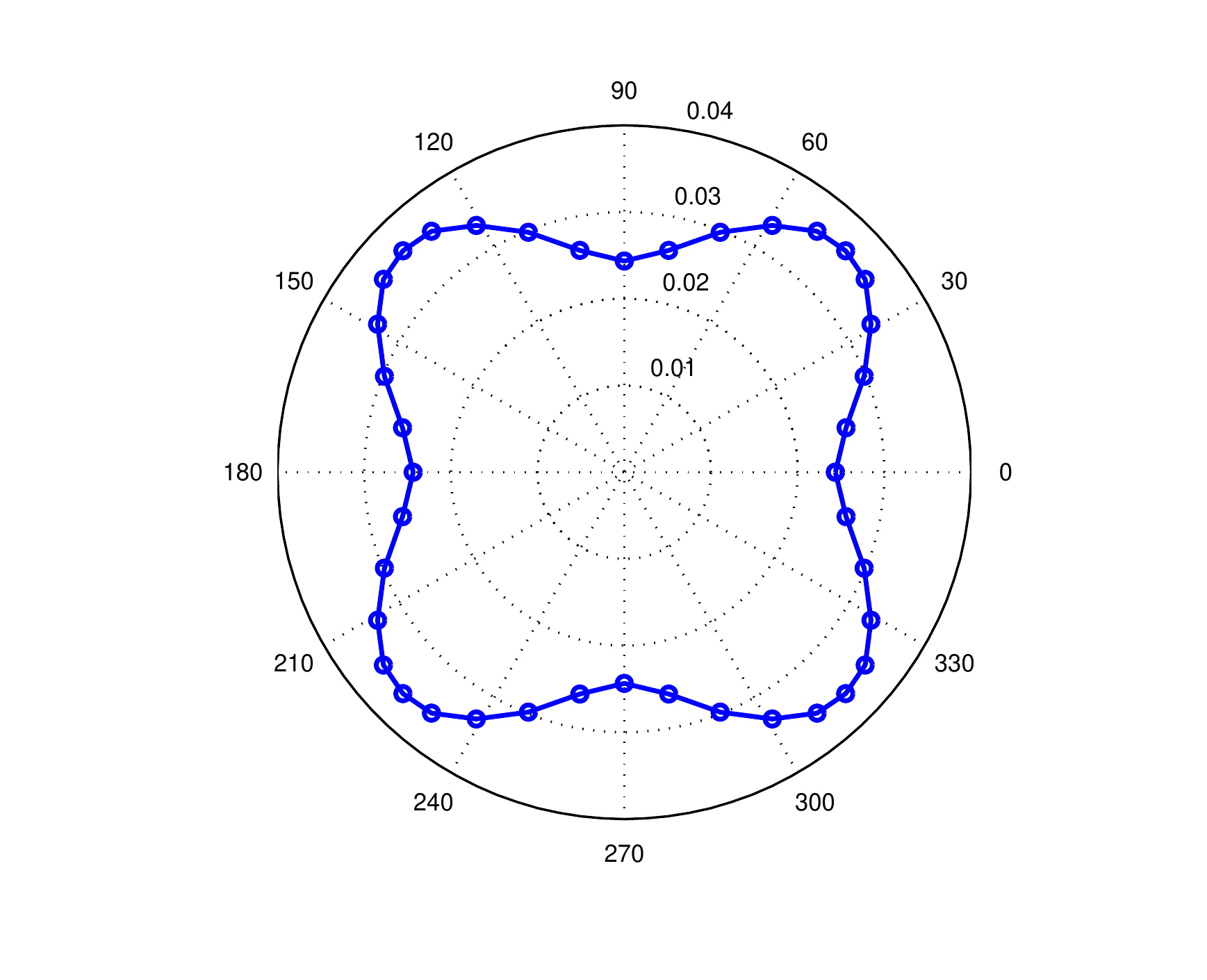} 
\caption{\label{fig18} Polar plot of 2D interfacial free energy (eV/\AA$^2$). The 0 deg and 45 deg correspond to [100] and [110] direction, respectively.}
\end{figure}

Since the continuum scale phase field computations are performed on a 2D domain, the interfacial energy anisotropy within the plane (001) is considered, and the long axis is directed along [110]. The resulting polar plot of the interfacial energy at 800~K as calculated in the 10$\times$5$\times$5 system is shown in Figure~\ref{fig18}.

\subsubsection{Lithium ion kinetics}

High lithium ion mobility, which in part determines the battery charge/discharge rate, is one of the major criteria for selecting promising lithium ion battery electrodes.\cite{Sebastian2003,anton2013} As most intercalation electrodes experience phase transformations during battery charge/discharge (lithiation/delithiation), lithium ion diffusion within the crystallites is determined not only by its mobility in the single phases but also across interfaces in the two-phase region. Despite its importance, only an ``apparent diffusion coefficient'' is usually measured for the two-phase region by fitting experimental data to equations derived from Fick's law of diffusion and its reliability is doubtful.\cite{Zhu2010} Numerical results with phase field models show that the error induced is large when the lithium composition reaches the two phase region, especially within the spinodal region.\cite{han2004} Attempts have been made to develop models that explicitly treat phase boundary movement, e.g.,the  moving boundary model,\cite{Funabiki1999} and the mixed control phase transformation model.\cite{Zhu2010}

Lithium ion transport in the electrodes may be either limited by diffusion in the stable phases or controlled by the mobility of the interface. It has been reported that the rate of lithium insertion into graphite during two-phase reactions is determined by diffusion in the stable phases.\cite{Funabiki1999} In Li$_{4}$Ti$_5$O$_{12}$, in contrast, the measured diffusion coefficient in the stable phases is much higher than that in the two-phase region, which suggests that phase boundary mobility is rate limiting during Li insertion or removal.\cite{Li2012} Analysis of lithium diffusion in vanadium oxide films also suggests that the kinetics of the two phase process is controlled by the phase boundary mobility.\cite{bae1996} For more complex cases, asymmetry of lithium ion mobility in stable phases leads to asymmetrical charge/discharge process\cite{Li2012} and pinning of phase boundary movement.\cite{choi1998} Despite this dependency, most continuum scale models assume a diffusion-limited process to simplify the problem.\cite{hong2006}

In the present work we systematically studied the mobility of the interface, and found that the mobility in the interface region is about three orders of magnitude higher than the $\alpha$ phase, indicating that the mobility is in fact diffusion limited.\cite{TonghuThesis} For the continuum scale computations in the solid-solution phase, we consider a constant mobility corresponding to the self diffusion coefficient value of 10$^{-5}$~cm$^2$/s, and as explained earlier, the mobility is assumed to be orientation independent.

\begin{figure}
\includegraphics[width=3.0in]{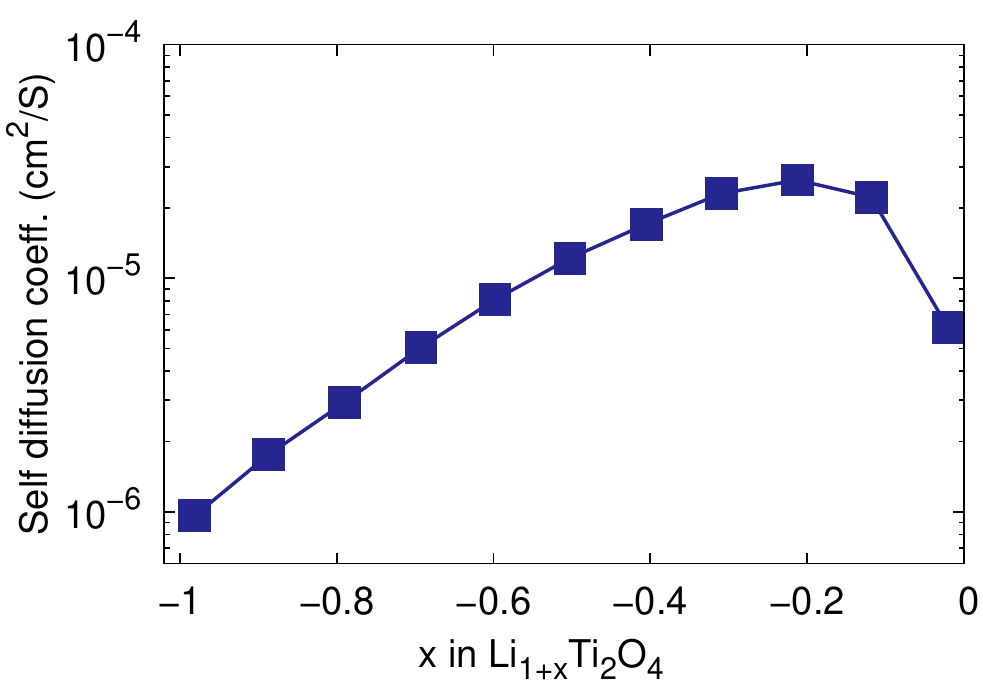} 
\caption{\label{fig:selfd} Self-diffusion coefficient as a function of composition in the $\alpha$ phase.}
\end{figure}

We calculated the lithium ion diffusion coefficient in the $\alpha$ solution phase ranging from $ x=-$1 to $x=$~0, and related it to the global mobility via Eq.~\ref{M_D_relation}. The kinetic Monte Carlo simulations of diffusion in the $\alpha$ phase using an energy landscape searching algorithm indicated that hops occurred exclusively between nearest neighbor tetrahedral sites; therefore only this type of transition was allowed in subsequent kinetic Monte Carlo simulations and the energy landscape searching algorithm was no longer used. Because the lithium diffusion coefficient in the solution phase of \lto\ is isotropic, a simple periodic boundary condition for studying diffusion of lithium in (100) was used, with a 6$\times$6$\times$6 unit cell at T=800~K. Straightforward KMC calculations with 500 to 1000 sweeps (in each KMC calculation, at each Li composition) were performed, and the corresponding diffusion coefficients were then calculated with the help of Eq.~\ref{selfd}. Figure~\ref{fig:selfd} shows the calculated mobility in the solid solution phase at T=800~K. The trend of the curves is the same as the diffusion coefficient calculated by Bhattacharya {\it et al.}\cite{Bhattacharya2010} The diffusion coefficient reported there was smaller as it was calculated at room temperature, in contrast to 800~K in the present work.

\subsection{Continuum scale simulations}

The continuum scale computations are primarily aimed at (a) understanding the evolution of lithium ion composition and diffusion kinetics during charging/discharging cycles, and studying the effects of heterogenous nucleation sites or defects on the overall diffusion kinetics, and (b) studying the mechanical response and stress variations during the diffusion processes due to the dependence of the elastic moduli and lattice parameters on the lithium ion composition. These two studies have been constructed so as to provide a qualitative picture of the efficiency of lithiation/delithiation processes in terms of charge distribution and localization, and also to understand the effect of spatial and temporal stress fluctuations on the material degradation that can affect the cycleability of these electrode particles. First, we describe the problem setup (geometry, initial conditions and boundary conditions), and then present the results of the continuum scale studies. 
 
\subsubsection{Problem setup}

\begin{figure}
\psfrag{a}{no nuclei}
%\psfrag{b}{one nucleus}
\psfrag{c}{two nuclei}
\psfrag{d}{three nuclei}
\psfrag{e}{five nuclei}
\includegraphics[width=2.5in]{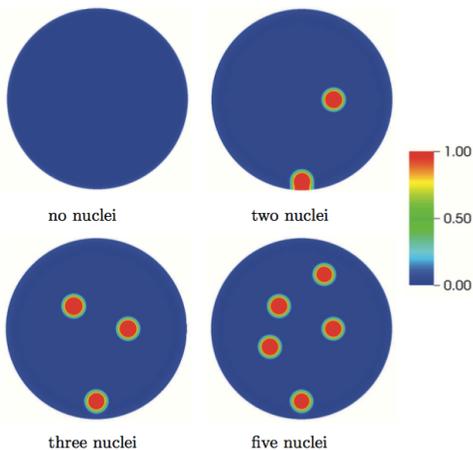} 
\caption{\label{fig:nucleationsites} Problem geometry (circular domain of radius \SI{1.0}{\micro\metre}) and composition initial conditions corresponding to homogenous nucleation (no nucleation sites) and heterogenous nucleation (2, 3 and 5 nucleation sites). Shown are the locations of the nucleation sites, each modeled as a circular region of \SI{0.1}{\micro\metre} radius with a composition value $x = 1$. The legend indicates the Lithium ion composition.}
\end{figure}

\begin{figure}[hbt]
%    \begin{tikzpicture}
 %     \begin{axis}[minor tick num=1,xlabel={Time (s)},ylabel={Flux ($s^{-1}\AA^{-2}$)},  xmin=0, xmax=0.001, legend entries={}, legend style={draw=none, at={(0.75,1)},anchor=north west}]
 %      	\addplot [black, thick] table [x expr=\thisrow{time}*0.001, y expr= \thisrow{flux}*10000000] {\cTwoNuclei};
 %     \end{axis}
 %   \end{tikzpicture}
  \includegraphics[width=2.5in]{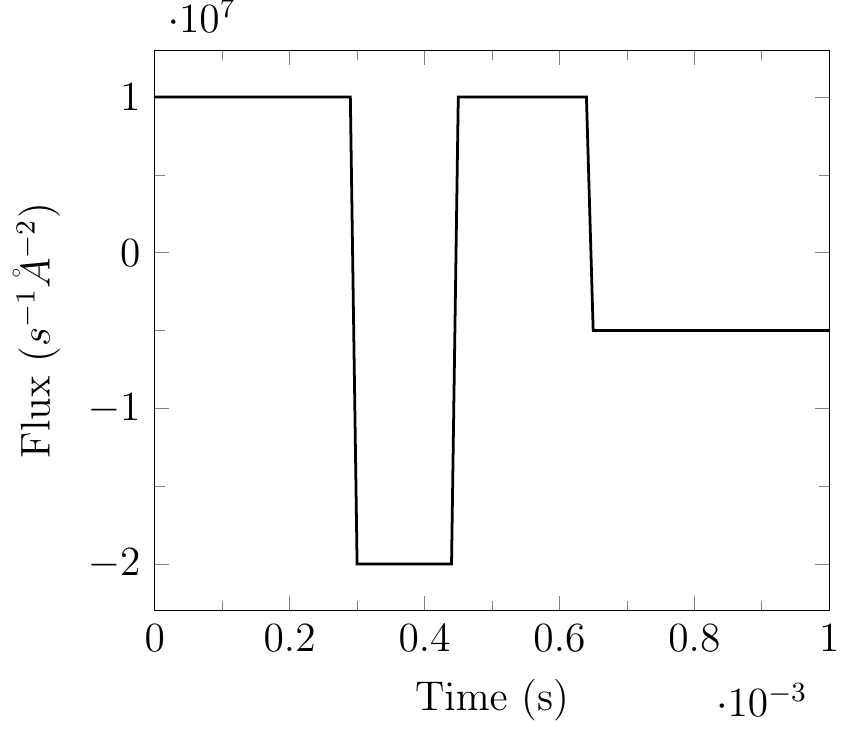} 
    \caption{\label{fig:fluxBC} The temporal variation of the Lithium ion flux applied at the boundary to model charging (lithiation) and discharging (delithiation) cycles. The flux magnitude considered here models an initial charging ($0 \le t < 0.3~\text{ms}$), followed by a fast discharging ($0.3~\text{ms} \le t < 0.45~\text{ms}$) and charging ($0.45~\text{ms} \le t < 0.65~\text{ms}$) and finally slow discharging ($0.65~\text{ms} \le t < 1.0~\text{ms}$).}
\end{figure}

We consider a circular domain of radius \SI{1.0}{\micro\metre} as the geometry of the electrode particle, and all the computations presented here are restricted to two dimensions (Figure~\ref{fig:nucleationsites}). Mechanically, the electrode particles are only subjected to a minimal set of Dirichlet boundary conditions on the displacement field to prevent rigid body motion. The composition field is driven by radially symmetric flux (Neumann) boundary conditions that induce an inward/outward flux of lithium ions. The temporal variation and cycling of the boundary flux (inward/outward) are depicted in Figure~\ref{fig:fluxBC}. These conditions on the displacement field and the Lithium ion flux remove any null space from this two-field (composition and displacement) multi-physics formulation and ensure the numerical stability of the simulations. Finally, since the computations are initial boundary value problems (IBVPs), one also needs to specify initial conditions on the composition field. We use the initial conditions to specify either a system in which homogenous nucleation dominates (no heterogeneous nucleation sites) or a system in which nucleation is dominated by the presence of nucleation sites, modeled here as circular regions of \SI{0.1}{\micro\metre} radius (Figure~\ref{fig:nucleationsites}). The motivation for including nucleation sites is driven by the physical understanding that these electrode particles are not spatially uniform and may have material defects or other heterogeneities, which enhance  nucleation. It is expected that heterogeneous nucleation dominates over homogeneous nucleation in all except defect-free systems. 

Recall that since the diffusion kinetics is very slow compared to elastic wave speeds, we assume the material to always be in elastic equilibrium and neglect inertial effects. Thus we model the mechanics problem as quasi-static without inertial terms. The initial conditions for the mechanics problem are zero displacement field with no pre-stress.
%data files
%\pgfplotstableread{data/simulations/noNuclei/data.dat}{\cNoNuclei}
%\pgfplotstableread{data/simulations/oneNuclei/data.dat}{\cOneNuclei}
%\pgfplotstableread{data/simulations/twoNuclei/data.dat}{\cTwoNuclei}
%\pgfplotstableread{data/simulations/threeNuclei/data.dat}{\cThreeNuclei}
%\pgfplotstableread{data/simulations/fiveNuclei/data.dat}{\cFiveNuclei}

\subsubsection{Lithium diffusion kinetics}
In this study, we consider four different IBVPs, each with different initial conditions used to model the homogenous/heterogenous nucleation. The first set of computations were carried out assuming a homogenous domain with no nucleation sites, and the subsequent computations considered multiple spatially distributed nucleation sites. Here, we only present results for the problems with two, three, and five nucleation sites whose sizes and spatial distributions are shown in Figure~\ref{fig:nucleationsites}. As previously stated, the lithiation/delithiation process is modelled by the application of an axisymmetric inward/outward flux on the surface of the circular domain. The time-varying flux profile considered in these studies is shown in Figure~\ref{fig:fluxBC}, where a positive value indicates influx and a negative value indicates outflux. 
  
\begin{figure}
\psfrag{a}[cc][][0.99][0]{\small t=0.3}
\psfrag{b}[cc][][0.99][0]{\small t=0.45}
\psfrag{c}[cc][][0.99][0]{\small t=0.6}
\psfrag{d}[cc][][0.99][0]{\small t=0.9}
\psfrag{e}[cc][][0.99][90]{\small no nuclei}
\psfrag{f}[cc][][0.99][90]{\small two nuclei}
\psfrag{g}[cc][][0.99][90]{\small five nuclei}
\includegraphics[width=3.5in]{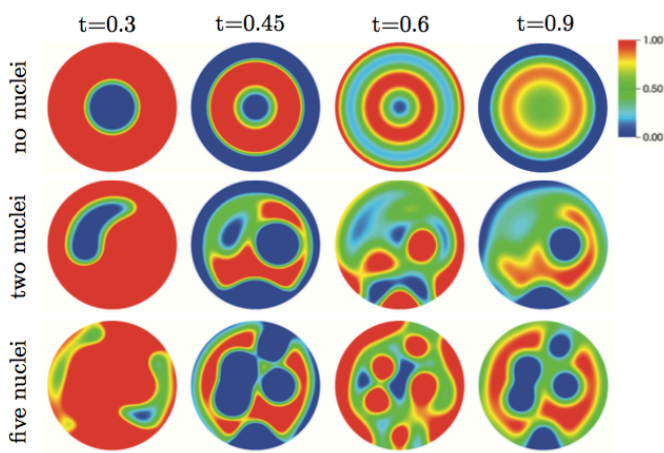} 
\caption{Time snapshots of Lithium ion composition during the lithiation/delithiation cycles. Shown are the composition fields at time $t=0.3~\text{ms}$, $t=0.45~\text{ms}$, $t=0.6~\text{ms}$ and $t=0.9~\text{ms}$ for the problems with no nuclei, two nuclei and five nuclei. The blue regions with composition close to $x=0$ are in the $\alpha$ phase and the red regions with composition close to $x=1$ are in the $\beta$ phase. Intermediate values of the composition correspond to the two-phase regions.}
\label{fig:composition}
\end{figure}

The temporal evolution of lithium composition during the lithiation/delithiation cycles shows patterns of phase evolution within the electrode particle. Figure~\ref{fig:composition} shows the time evolution of lithium ion composition in the presence of zero, two and five nucleation sites. With no nucleation sites there is no breaking of the axisymmetry of the composition field. However, even in this case, the rapid lithiation/delithiation leaves behind regions of depletion ($t=0.45~\text{ms}$) and accumulation ($t=0.9~\text{ms}$). This suggests that to avoid such phase localization within electrode particles, the lithiation/delithiation rates must be carefully controlled. These patterns of lithium localization are more prominent as well as complex in computations with heterogeneous nucleation sites, which break the symmetry. In these cases, two important observations are that (i) lithium depletion/accumulation occurs primarily around  nucleation sites, and that (ii) heterogeneous nucleation leads to complex spatial patterns of lithium distribution as seen in the two nuclei and five nuclei case at time $t=0.3~\text{ms}$, $t=0.45~\text{ms}$, $t=0.6~\text{ms}$ and $t=0.9~\text{ms}$. 

\subsubsection{Mechanical response and stress evolution}
\begin{figure}
\psfrag{a}[cc][][0.99][0]{\small t=0.3}
\psfrag{b}[cc][][0.99][0]{\small t=0.45}
\psfrag{c}[cc][][0.99][0]{\small t=0.6}
\psfrag{d}[cc][][0.99][0]{\small t=0.9}
\psfrag{x}[cc][][0.99][0]{\small 1.6}
\psfrag{y}[cc][][0.99][0]{\small 0.0}
\psfrag{z}[cc][][0.99][0]{\small -1.6}
\psfrag{e}[cc][][0.99][90]{\small no nuclei}
\psfrag{f}[cc][][0.99][90]{\small two nuclei}
\psfrag{g}[cc][][0.99][90]{\small five nuclei}
\includegraphics[width=3.5in]{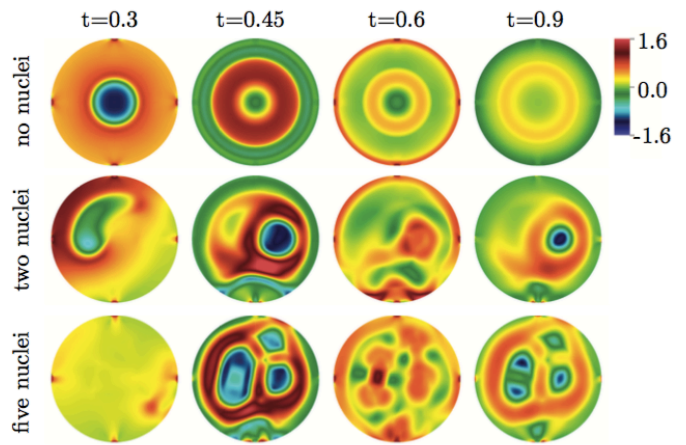} 
\caption{Time snapshots of maximum principal stress distribution during the lithiation/delithiation cycles. Shown are the stress fields at time $t=0.3~\text{ms}$, $t=0.45~\text{ms}$, $t=0.6~\text{ms}$ and $t=0.9~\text{ms}$ for the problems with no nuclei, two nuclei and five nuclei. The unit of stress in the contour plots is GPa.}
\label{fig:stress}
\end{figure}

An important factor in the performance degradation of electrode materials is the mechanical failure of electrode particles through fracture\cite{McMeeking2016, Miehe2016}. While we do not explicitly model material degradation by fracture/damage mechanisms in the current work, we can still obtain a useful qualitative picture of material degradation by tracking the stress distribution and evolution of peak stress values during the lithiation/delithiation cycles. The peak stress values are important, since fracture and voiding in brittle and ductile materials, respectively, are initiated at points with critical values of the relevant stress measures such as the maximum principal stress and the related Mode-I/Mode-II opening tractions for fracture, and the maximum hydrostatic stress for voiding. In this work we identify the temporal evolution of the maximum principle stress values from which the other measures can be obtained.  The spatial distribution of stress and stress localization observed in the presence of zero, two and five nucleation sites are shown in Figure~\ref{fig:stress}. 

The computations indicate that the peak stress values occur at the surfaces of the nucleation sites and at the sites of lithium localization. This spatial distribution of peak stresses is explained by the fact that the interfaces are phase boundaries characterized by discontinuously changing lattice parameters and elastic moduli, which  induce large stresses. Further, the spatial distribution of peak stress values is strongly dependent on the heterogeneity induced by the nucleation sites. As seen in Figure~\ref{fig:stress}, the peak stress profiles are significantly different for the IBVPs with two and five heterogeneous nucleation sites when compared to the case with no heterogeneous nucleation sites.

\begin{figure}[hbt]
%    \begin{tikzpicture}[scale=1.25]
 %     \begin{axis}[minor tick num=0,xlabel={Time (s)},ylabel={Flux ($s^{-1}\AA^{-2}$)},  legend entries={Flux}, legend style={draw=none, at={(0.75,1)},anchor=north west}]
 %      	\addplot [dashed, black, thick] table [x expr=\thisrow{time}*0.001, y expr= \thisrow{flux}*10000000] {\cTwoNuclei};
 %     \end{axis}
 %     \begin{axis}[minor tick num=0, axis y line*=right, axis x line=none, ylabel={\text{Stress (GPa)}}, legend entries={0, 2, 3, 5}, legend style={draw=none, at={(0.75,.93)},anchor=north west}]
 %      \addplot [gray, thick] table [x expr=\thisrow{time}*0.001, y expr= \thisrow{stress}*160] {\cNoNuclei};
       %\addplot [blue, thick] table [x={time}, y={stress}] {\cOneNuclei};
 %      \addplot [green, thick] table [x expr=\thisrow{time}*0.001, y expr= \thisrow{stress}*160] {\cTwoNuclei};
 %       \addplot [cyan, thick] table [x expr=\thisrow{time}*0.001, y expr= \thisrow{stress}*160] {\cThreeNuclei};
 %       \addplot [red, thick] table [x expr=\thisrow{time}*0.001, y expr= \thisrow{stress}*160] {\cFiveNuclei};
 %       \end{axis}
 %   \end{tikzpicture}
    \includegraphics[width=3.0in]{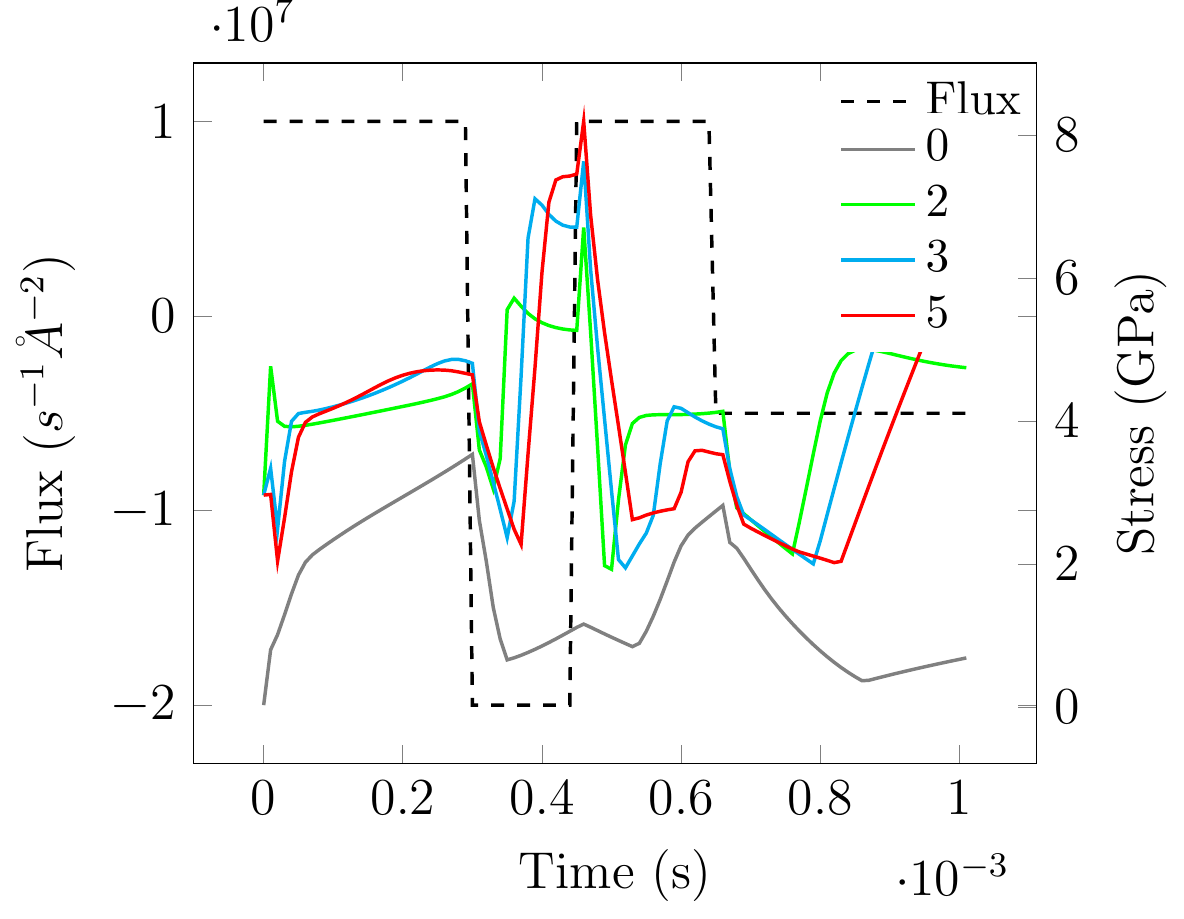} 
    \caption{Variations in the peak values of the maximum principal stress during the lithiation/delithiation cycles in the presence of 0, 2, 3 and 5 nucleation sites. The flux variation indicating lithiation/delithiation cycles is superposed for reference.}
    \label{fig:stressWithTime}	
\end{figure}

Using a maximum principle stress criterion for fracture, one can expect to see crack initiation and evolution from material points in the domain where the peak maximum principle stress value exceeds a critical value. In the absence of data about the critical value of the maximum principle stress for crack initiation, we examine the evolution of peak stress values in the electrode particle and the effect of lithiation/delithiation cycles by plotting the time evolution of the peak maximum principle stress magnitude (Figure~\ref{fig:stressWithTime}). These simulations indicate that the peak stress values are higher in the case of heterogenous nucleation and that the peak values increase with the number of nucleation sites. This observation, coupled with the spatial distribution of stress shown in Figure~\ref{fig:stress} indicate that there is an enhanced possibility of fracture and voiding in the vicinity of nucleation sites.

Further, we see a convergent behavior with respect to number of nucleation sites, clearly shown by the similar peak stress profiles for the cases of two, three and five heterogeneous nucleation sites. This indicates that in the limit of numerous randomly located heterogeneous nucleation sites, one could obtain a characteristic peak stress distribution profile for a given electrode particle geometry and lithiation/delithiation cycles. Some details of our observations, such as the stress profile and its time evolution, may be influenced strongly by the specific lithiation/delithiation cycles chosen in these simulations, the circular geometry, and other IBVP-specific choices.  However, the broad qualitative trends of the peak stress profiles and their time evolution can be very important to the design of lithiation/delithiation cycles and the understanding of the electrode particle performance and possible degradation. 

\section{Conclusion}
The atomistic calculations on the \lto\ system reported here add to the calculations reported in Ref.~\citenum{Bhattacharya2010}, and supply the necessary quantities used in the continuum scale studies. Here we have reported the elastic coefficients C$_{11}$, C$_{12}$ and C$_{44}$ of \lto\ , as well as the Gibbs free energy and the interfacial energy between the $\alpha$ and the $\beta$ phases of \lto\ at 800~K. The calculated mobility indicates that transport in \lto\ is diffusion-limited. This same approach could be implemented, and perhaps automated, for a wide array of electrode materials in order to provide the groundwork for analogous simulation studies. This may encourage wider application of these methods to study battery electrode materials.
 
For the continuum scale studies, we considered a phase field framework for modeling the chemo-mechanical problem of phase transformations coupled with mechanics. The numerical framework is based on a coupled phase field and finite strain mechanics formulation. The latter choice distinguishes this work as traditionally phase field computations are limited by the infinitesimal strain approximation.

The continuum scale simulations provide insights to the kinetics (spatial distribution of Lithium ion composition) and mechanics (stress profiles and peak stress value evolution) and how these are affected by the lithiation/delithiation processes.  Specifically, the peak stress profile is observed to be associated with the presence and the density of nucleation sites and its time evolution is associated with the rates of the imposed lithiation/delithiation cycles. As discussed, these stress profiles and peak stress values are important for understanding fracture and voiding.

Thus, continuum scale studies, informed by first principles,  MMC and KMC studies, have a potential role to play in optimizing lithiation/delithiation cycles and predicting the effects of nucleation sites on the overall performance and mechanical degradation of electrode particles. First principles parameterizations of continuum simulations allow us to better understand battery performance via the simulation of lithiation/delithiation dynamics and studying the resulting Lithium ion diffusion kinetics and mechanical response, and thus potentially contribute to the design optimization of Lithium ion batteries.

\section*{Acknowledgement}
This research was supported by the NSF CDI Type I grant: CHE1027729 ``Meta-Codes for Computational Kinetics''. First principles calculations were performed on the Homewood High Performance Computing Cluster supported in part by the U.S. National Science Foundation, grant NSF-OCI-108849. The continuum scale computations were performed on a PRedictive Integrated Structural Materials Science (PRISMS) Center computing cluster at the University of Michigan. PRISMS is supported by the U.S. Department of Energy, Office of Basic Energy Sciences, Division of Materials Sciences and Engineering under Award \#DE-SC0008637. AR was supported by NSF DUE1237992. 
\bibliography{paper}

\end{document}